# A Data-Centric Methodology and Task Typology for Time-Stamped Event Sequences

Yasara Peiris*†  Clara-Maria Barth*‡  Elaine M. Huang§  Jürgen Bernard¶
University of Zurich    University of Zurich    University of Zurich    University of Zurich
                                                                        Digital Society Initiative

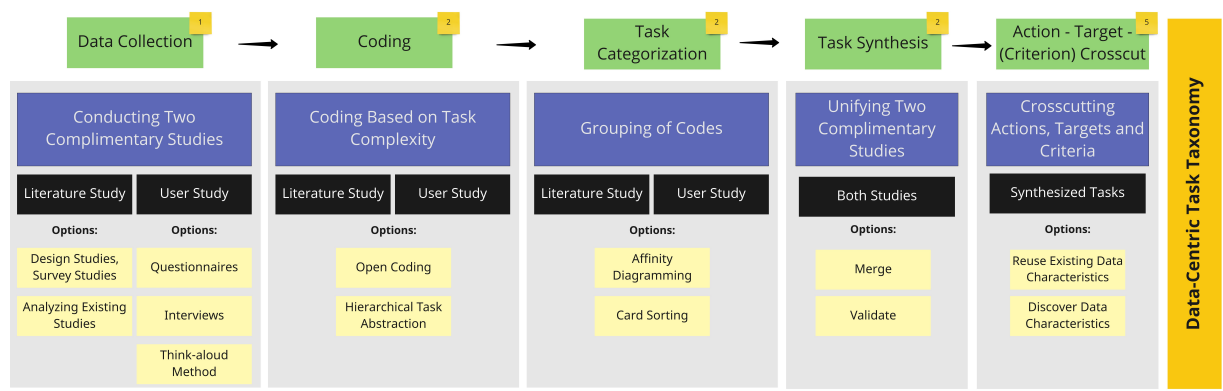

Figure 1: Our methodology for task abstraction and taxonomy-building for dataset-specific tasks comprises five phases. A variety of methods can be applied in the three early phases of data collection, coding, and task categorization, for moving forward with two study sources. Phase four includes a task synthesis, followed by the fine-grained elaboration on action-target-(criterion) crosscuts.

**ABSTRACT**

Task abstractions and taxonomic structures for tasks are useful for designers of interactive data analysis approaches, serving as design targets and evaluation criteria alike. For individual data types, dataset-specific taxonomic structures capture unique data characteristics, while being generalizable across application domains. The creation of dataset-centric but domain-agnostic taxonomic structures is difficult, especially if best practices for a focused data type are still missing, observing experts is not feasible, and means for reflection and generalization are scarce. We discovered this need for methodological support when working with time-stamped event sequences, a datatype that has not yet been fully systematically studied in visualization research. To address this shortcoming, we present a methodology that enables researchers to abstract tasks and build dataset-centric taxonomic structures in five phases (data collection, coding, task categorization, task synthesis, and action-target-(criterion) crosscut). We validate the methodology by applying it to time-stamped event sequences and present a task typology that uses triples as a novel language of description for tasks: (1) action, (2) data target, and (3) data criterion. We further evaluate the descriptive power of the typology with a real-world case on cybersecurity.

## 1 INTRODUCTION

The abstraction of analysis tasks is a useful method that allows visualization designers to build tools to support users in specific applications and problem contexts [5, 56, 58, 68, 74]. Task abstractions help to address individual information needs and datasets that contain special characteristics. Task abstractions form design targets for interactive visual data analysis approaches, and can further be used as a basis for iterative or summative validations. Task taxonomies, classifications, and typologies are complete, disjoint, and systematically structured organizations of tasks in a focused analysis context, such as for specific data types; we will use the term *taxonomic structures* when referring to these types of task organizations. They are the result of complex thought processes, such as reflections on existing approaches, interviews with users, and expert knowledge of researchers. Taxonomic structures for tasks enable researchers and visualization designers to benefit from a) an easy way to reason about complex matters, b) comparability across existing approaches, and c) guidance for the design of novel solutions.

This work is inspired by a concrete data type with surprisingly little exposure in the visualization community: time-stamped event sequences, in short: *TSES*, for singular and plural. A TSES is a sequence of time-stamped events with temporal information, but *no* value information. Thus, the primary analysis target is only the temporal characteristics of event signatures. This is in contrast to other time-oriented data types [2] such as classical event sequences (categorical values), univariate (numeric values), and multivariate time series (multiple values per time stamp), Figure 2 illustrates these differences. TSES are relevant in almost any application domain, including manufacturing, healthcare, education, or geology. Example cases of TSES include git commits, hospitalizations due to Covid-19, customer complaints, traffic accidents, sleep behavior, or earthquake aftershocks. For the sense-making process, phenomena encoded with TSES are often accompanied by rich sets of data attributes (metadata), such as the sensor device used for a measurement, the cohort of a patient record that was studied, or the geographic region of some Earth phenomenon. Such metadata attributes can be used to contextualize observed phenomena, such as the day of the week explaining observed sleep behavior.

For the identification of tasks and taxonomic task structures for TSES, we identify four main challenge areas. First, TSES are a complex and specific data type with special analysis tasks. TSES have three natural granularities: events, event sequences, and groups of event sequences, with metadata at each granularity. Furthermore, in data mining research, TSES are often assessed with metrics which produce features to support the analysis, such as peaks, gaps, recurrences, accelerations, or outliers. Second, visualization designers have limited opportunities to learn from existing best practices on TSES, as few visualization approaches and design studies exist for TSES [20, 50, 61, 86, 91]. Third, no data-specific taxonomic structure for tasks exists for TSES, in contrast to other time-oriented data types. In fact, a series of datatype-agnostic taxonomic structures exist, but are often too general for the specific set of dataset-centric tasks. Datatype-specific and less generic task descriptions can be beneficial [14], as several best practice examples show [1, 2, 33, 45].

*shared first authorship
†e-mail: yasara.peiris@uzh.ch
‡e-mail: clara-maria.barth@uzh.ch
§e-mail: huang@ifi.uzh.ch
¶e-mail: bernard@ifi.uzh.ch

| Data Type | Values | Time | Example |
|---|---|---|---|
| Time-stamped event sequences | none | 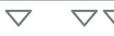 | Heartbeats |
| Classical event sequences | 1x cat | 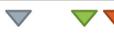 | Treatment plan |
| Univariate time series | 1x num | 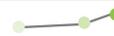 | Stock chart |
| Multivariate time series | nx num/cat | 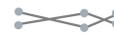 | Sensor data |

Figure 2: Characterization of TSES and differentiation from other time-oriented data. TSES have only temporal signatures but no values.

Finally, we identify the need for methodological support for taxonomy-building for specific data types, especially when few best-practice approaches exist and reflecting on limited cases could yield subjective rather than empirical and generalizable outcomes. Borrowing task abstractions from related data types is inspiring [41], but may lack precision and completeness. Observational studies, think-aloud methods, and contextual inquiry require the active use of existing visualizations and thus suffer from bootstrap problems. Interviews and surveys conducted with large numbers of potential users can yield insights for task generation [3], but may differ from analysis of problems by domain experts. In summary, for less investigated data types such as TSES, existing methodologies are not ideal for building data-centric taxonomic structures. Our main concern is whether using only a single methodology is useful.

To arrive at a data-specific and domain-agnostic typology for TSES given these four challenges, we propose four contributions:

1. A data-specific and domain-agnostic methodology for abstracting tasks and building taxonomic structures, combining two complementary sources of information: the study of literature and the study of users. Information from both is systematically processed, aligned, and synthesized in five steps.
2. We validate the methodology by applying it to TSES. Overall, we examined (A) 65 surveys of participants and (B) 16 design studies, each source relevant to TSES, spanning various domains, and covering a highly diverse set of TSES analysis cases.
3. We report on the notation of tasks using triples of (1) an *action* that is applied to (2) a *target* (data), by the use of a (3) *criterion* (data). This extension of tuples (e.g., actions and targets) substantiates the description of abstract actions through concrete targets and criteria, for data-specific taxonomic structures.
4. We propose a data-centric and domain-agnostic typology of analysis tasks for TSES to guide the design of novel visualization approaches. We validate the descriptive power of the typology by characterizing a concrete real-world case for TSES.

Our methodology offers advantages for applications in uncharted territory and for data types that have not been extensively investigated. Furthermore, a multi-strand approach using complementary studies reduces the risk of misclassifying or overlooking tasks [41]. With the description of tasks based on triples, we open the space for more fine-grained, specific, and dataset-centric descriptions of tasks, which may inspire future taxonomic approaches beyond TSES.

## 2 RELATED WORK

We review specifics of TSES and associated tasks, gaps of taxonomic work for TSES, and methodologies for taxonomy-building.

### 2.1 Tasks for Time-Stamped Event Sequences

TSES are temporal data without explicit values. In contrast, most related approaches for time-oriented data focus on classical event sequences with categorical events (such as the treatment types for a patient) or numerical time series, either univariate (a stock chart) or multivariate (multiple sensors) (cf. Figure 2). A survey of classical event sequences by Guo et al. [33] categorizes approaches by the application domain and analysis tasks. In the medical domain, analysis tasks include sequential pattern mining [32, 65, 77], clustering sequences [28], comparing sequences [29, 38, 87], and comparing patient cohorts [51]. Similarly, in the domains of web data exploration and clickstream analysis, analysis tasks often comprise pattern mining [46, 84], aggregation/clustering [85, 89], comparison of multivariate sequences [95], and detecting anomalies [90]. Other application domains require the analysis of user behavior using internet data [15, 63, 94], detecting anomalies in manufacturing [91], and predictive analysis and recommendations [18, 19, 30].

One difference between TSES and other time-oriented data is the need for metadata attributes, allowing users to contextualize findings. Examples include information about the experiment, device names, patient demographics, or geographic regions. This aligns TSES with relate [75] or relation-seeking [5] tasks, in which an interesting pattern/finding in a data subset is contextualized and associated with explanatory attributes or metadata [6, 9–11]. Finally, analysis tasks align with metrics that can be applied to TSES [52, 60] for the identification of features in the temporal event signature, such as regularity, acceleration, gaps, entropy, or spikes [25, 27, 43, 79]. In summary, TSES include specific characteristics, leading to various specific analysis tasks. However, despite interesting anecdotal insight into tasks, no taxonomic structures for TSES exist and most task abstractions in related design studies have been done manually [20, 50, 86].

### 2.2 Task Taxonomies, Classifications, and Typologies

We review taxonomic structures according to task characteristics that have shaped our work: the degree of abstraction and the granularity.

The *abstraction* of (data and) tasks is a fundamental concept for expressing special or unique entities in a language that makes them comparable. Abstracted tasks allow the systematic study, the elaboration of design spaces, the reuse of existing visualization methods, and the validation of visualization designs [5, 59, 68]. The abstraction from the problem domain is applied in many visualization approaches, guided by design frameworks such as the data-user-task triangle [56] or the nested model [58]. In contrast, domain-specific taxonomies explicitly focus on domain context, e.g., in comparative genomics [55] or network security [78]. The abstraction from data specifics leads to data-agnostic taxonomies that are generalizable, but often too broad for dataset-centric design approaches. Examples of data-agnostic taxonomies, classifications, and typologies include proposals by Wehrend and Lewis [88], Amar and Stasko [4], Roth [69], Thomas and Cook [81], Natalia and Gennady Andrienko [5], Yi et al. [92], Schulz [71], and Brehmer and Munzner [13]. In turn, specific data types can benefit from less generic task descriptions [14], at the expense of generalizability. Examples exist for dimensionally-reduced data [14], temporal networks [1], classical event sequences [33], time series [2, 76], or graphs [45]. However, no taxonomic structures have been proposed for TSES so far.

The *granularity* of task descriptions varies considerably, from complex and diverse to fine-grained and atomic representations. Rind et al. [68] present an ordering of task granularities and corresponding terminology from goals to activities, to tasks/sub-tasks, to actions, and operations. Higher-level and lower-level tasks form hierarchical relations, and it is widely accepted that complex goals can be broken down into more actionable tasks. Examples of higher-level tasks [68] are taxonomic structures by Amar and Stasko [4], Roth [69], and Thomas and Cook [81]. In contrast, low-level taxonomic structures typically focus more on atomic user actions and often align with interaction taxonomies, such as the categorization by Yi et al. on user intents [92]. For our task typology on TSES, we consider the low-level to mid-level granularity to be the most useful, sufficiently fine-grained that each task is its own actionable operation but coarse enough to differ from atomic (inter-)actions. Related mid-level taxonomic structures are presented by Wehrend and Lewis [88], Amar et al. [3], Pretorius et al. [67], and Natalia and Gennady Andrienko [5]. Taxonomic structures can also be borrowed from the data mining community, as Fayyad et al.'s work on knowledge discovery in databases shows [23]. A useful concept, we incorporated into the structuring of our typology was Munzner's subdivision of tasks into *actions* and *targets* [59], in which actions refer to what users (want to) do, and targets refer to the data focus at hand. We extend this notion and propose a triple-syntax for TSES.

## 2.3 Building Taxonomic Structures of Analysis Tasks

A prominent source of information for taxonomy-building is existing literature, including existing systems, techniques, problems, or user evaluations [44]. Such reflections require similar literature [41], which may not necessarily exist for a specific focus such as on TSES. Observational studies [37, 40, 82], think-aloud methods [8], and contextual inquiry [74] are useful methods for identifying tasks while using visualization systems, presuming that a solution for a problem already exists. In particular, domain experts can be a valuable source for task abstractions, through interviews [21, 22, 39, 73], surveys [3, 7] or other inquiries. Shortcomings include the limited availability of experts [70] and the risk of skewing the task set because of a low number of experts [42]. Data-first design studies [64] put real-world datasets (not necessarily the data type in general) into focus, which can be useful for defining the problem domain and for data-centric abstractions, even if the low number of stakeholders involved may reduce generalizability. Methods to arrive at abstracted tasks through design study [54] include Munzner's nested model [58], or hierarchical task abstractions [93], similar to our approach with the survey study. Finally, generalization efforts made in design studies [58, 74] allow a transfer of solutions to other problems or datasets [72]. For this reason, we apply existing design studies and application-driven work to related data types as one source for abstractions.

Building taxonomic structures benefits from methods to organize ideas and data, including iterative and focused coding [14, 16, 26], affinity diagramming [3, 34, 36, 92], and card sorting with domain experts [69]. Similar to our coding approach, Lam et al. [44] surveyed design studies using an open-coding process and created affinity diagrams to develop a task taxonomy. Finally, we drew inspiration from methods synthesizing existing task taxonomies to derive novel taxonomic structures [1, 13, 71, 92]. This inspired us to consider a triangulation-based approach to constructing a task classification by synthesizing task abstractions from two sources: user surveys and task abstractions from design studies in our applied case. In summary, downsides of individual methodologies exist, but can be effectively mitigated by combinining approaches [41].

## 3 METHODOLOGY

We propose a novel methodology for building dataset-specific and domain-agnostic taxonomic structures for tasks, depicted in Figure 1. Methodologies provided in the visualization literature are inspiring, however, it is a challenge when only applying a single methodological approach to a data type that has been little investigated (cf. Section 2.3). To close this gap, we follow a multi-strand approach to task gathering [42] and taxonomy-building, i.e., two complementing sources are acquired to arrive at a taxonomic structure. Sources of information for this methodological approach are users, e.g., through interviews, surveys, or observational studies, and the visualization literature, e.g., through surveys or abstractions derived from design studies. We propose a 5-stage workflow, consisting of three phases with task abstractions for the two sources, a task synthesis phase, and an action-target crosscut. We use the term *researchers* to refer to visualization researchers conducting the proposed methodology. To make our methodology actionable for researchers, we will apply our methodology to TSES in Section 5.

### 3.1 Data Collection

Researchers start the triangulation approach by gathering data from two study sources with complementary bodies of knowledge. User data can be collected using surveys [3], think-aloud sessions [8], contextual inquiry [74], email interviews [21], or semi-structured interviews [39] with domain experts. The data can cover a variety of information needs, analysis challenges, and analysis tasks. For many study participants, this approach can yield a long list of tasks while minimizing the risk of overlooking individual tasks. In turn, study data derived from visualization literature offers an overview of real-world problems, existing solutions, and an ample set of best practices from which to draw. Promising strategies for literature studies are surveys, taxonomic structures, and abstraction work applied directly to visualization literature [44, 72].

### 3.2 Coding

We suggest using coding [16, 26], a well-known early phase approach in qualitative research, with the aim of synthesizing concepts from the data collected. Examples of different codes are depicted in the neighboring figure and in the supplemental materials in detail. In this iterative process, it may be useful if researchers clearly differentiate 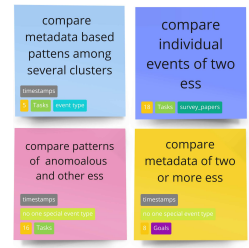
between high-level task descriptions and lower-level tasks to avoid confusion. Similarly, tasks with different levels of abstraction may receive individual treatment, to arrive at a more consistent description, e.g., to decompose complex tasks into smaller, more manageable tasks [68]. A practical guideline is provided by Zhang et al., who pioneered hierarchical task abstraction [93].

For lower-level task descriptions, open coding can be quite useful [82] as a means to identify the objectives, actions, and components while employing sensitizing questions [83]. An example of coding in visualization research is the work by Brehmer et al.s [14]. A final remark concerns the nature of data-centric tasks that typically combine an action and a target [71]. We observed that actions of tasks appear to be very descriptive, whereas targets may vary between data specifics, such as a comparison action that may be applied to different data targets including data items, or item groups (elementary vs. synoptic item granularity [5]). This is why it may be useful to focus on actions in the open coding process and investigate types of targets downstream to break down the process of taxonomy-building into more manageable pieces. Finally, when conducting open coding, it is advisable that two or more researchers assign codes independently. researchers then decide on a condensed code set based on code agreement, and by examining discrepancies between differing code assignments.

### 3.3 Task Categorization

The goal of this phase is to group the coded study data into more abstract tasks and to identify common tasks among the two sources. This can be achieved by affinity diagramming [3, 34, 92], and card sorting [69], and we refer to the work of Lam et al. for additional guidance on building affinity diagrams in visualization research [44]. We argue that researchers benefit from creating two task categorizations, for codes from the two sources. Two or more researchers conduct the grouping of codes iteratively, aiming at identifying unique tasks, as shown in the neighboring figure. Again, it may be useful to organize codes primarily by the actions expressed, with the tasks and focus on data targets downstream to reduce the complexity of the categorization challenge. The result of the task categorization phase is two task categorizations representing the abstracted tasks derived from the two sources, as it can be seen in our application in Section 5.3.

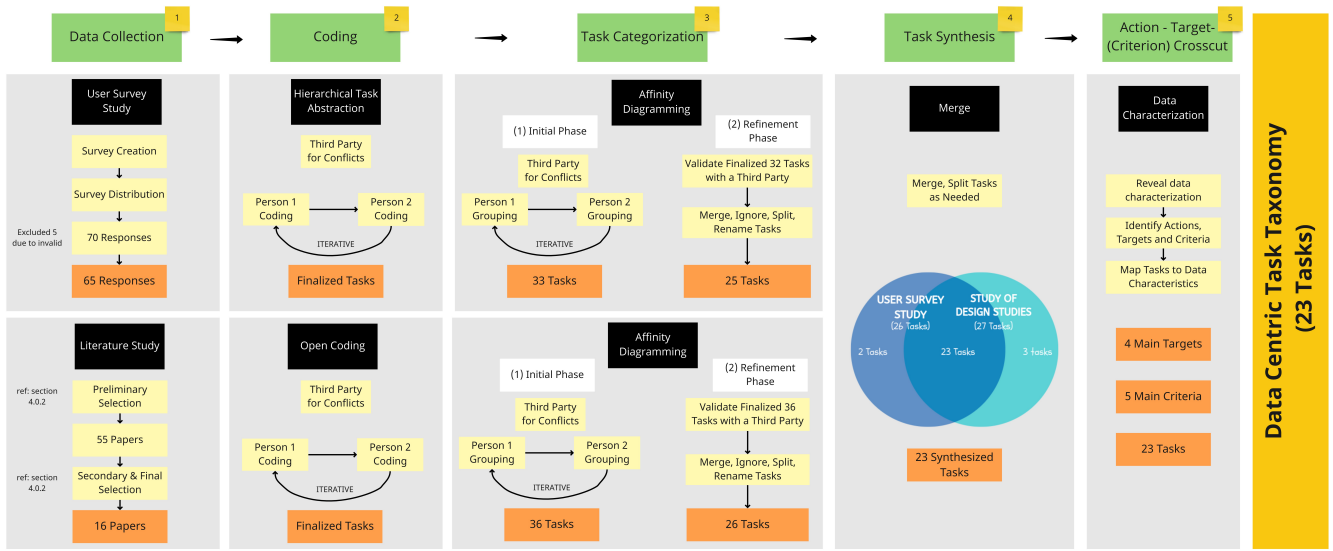

Figure 3: Application of our methodology for building a taxonomic structure using the case of TSES. For each of the five phases, we describe the concrete methods applied throughout. The first three phases handled two sources in parallel, followed by a synthesis phase, leading to 23 tasks. Finally, in the action-target-(criterion) crosscut, we identified four targets and five criteria, leading to the task typology presented in Section 6.

### 3.4 Task Synthesis

The task synthesis unifies the previously two-sourced approaches by joining the two spaces of abstracted task categorizations. This join can be achieved using two strategies, depending on the researchers' goal; (1) merging or (2) validating one by the other.

(1) When using the *merging* strategy, researchers treat the two task categorizations as equal partners, and both are used to form a unified set of tasks. To combine the task spaces represented with task categorizations, we suggest that researchers first identify and merge common tasks in both abstractions, before continuing with more heterogeneous tasks. If a merged task contains a high number of codes, researchers may consider a further subdivision to better represent individual nuances. In merging, the process naturally yields three subsets; (i) tasks found only in the user-based source, (ii) tasks found only in the visualization literature, and (iii) tasks that exist in both sources. Different strategies exist to manage subsets (i) and (ii). While inclusive task handling may preserve all tasks, an exclusive strategy may neglect (i), (ii), or both, leading to a more comprehensive taxonomic structure.

(2) If researchers have one primary source at hand, they may choose to *validate* the primary task categorization with the coding results of the second source. Different validation strategies can be applied to leverage the second source. One strategy is to assess the frequencies of codes per tasks to validate task relevance. For example, tasks that only exist in the primary source may be excluded, as they cannot be confirmed by the secondary source. Validation strategies may necessitate that different researchers develop the two alternative diagrams independently in the task categorization phase such that the integrity of the validation strategy is not compromised.

### 3.5 Action - Target - (Criterion) Crosscut

Finally, researchers put (1) *actions* of tasks into the context of the data (2) *targets* [59, 71] by generating a crosscut of actions and targets. It may be useful to further extend these tuples to triples to arrive at a more expressive description of tasks and included data characteristics. We present this notation of triples in Section 3.5 as a secondary contribution. We recommend performing this action-target-(criterion) crosscut as the final phase in the process, as a) researchers should have gained a clear vocabulary of actions to perform the crosscut effectively and b) separating an action-based focus from an action-target crosscut can reduce the complexity of the categorization and synthesis process. In this methodology, we assume that a general data characterization for the targeted data type exists. Examples of sources are the results of data abstraction [58] processes, and surveys of data-centric analyses, such as for time-oriented data [2, 24, 49, 57] or classical event sequences [33]. In fact, deriving a data characterization throughout the process is also feasible, with similar coding, categorization, and synthesis phases. The result of the action-target crosscut is a dataset-centric taxonomic structure for tasks as a baseline for precise and comprehensive recipes for building interactive visual data analysis approaches.

## 4 A Triple Notation: Action, Target and Criterion

We propose a novel way to represent tasks by making use of a triple (action, target, and criterion). The notion of triples builds upon and extends the state-of-the-art which often uses actions and targets for the characterization of abstracted tasks [59, 71]. *Actions* are usually represented by a verb and describe *why* users require visualization support to achieve their goals. *Targets* refer to *what* type of data characteristics the visualization pertains to, i.e., to what data the action is applied. Data and data characteristics may, e.g., differ in a) granularity (single items vs. item groups), b) focus on items or attributes/features (rows vs. columns in a tabular dataset), or c) the data content vs. metadata attributes. In the case of time-oriented data, it may also be useful to differentiate between a) points in time (local, events/timestamps) vs. entire time-oriented series (global, event sequence/time series), and b) temporal references (time) and value characteristics (values, temporal features, temporal metadata) [2, 5]. In addition to actions and targets, we introduce *criteria* as a third cornerstone to describe tasks. Like targets, criteria are typically a data characteristics. In contrast to targets, criteria describe the means to perform an action. As such, triples are inspired by Bertin's [12] and Andrienko and Andrienko's [5] notion of *targets* and *constraints*, where targets (the unknown, the question type) are to be identified through constraints (known information, our criteria). The benefit of triples is exemplified in the following example: "rank TSES by feature". Here, rank is the action performed, TSES is the target the action is applied on, features are the criteria by which the action is applied to the target. We discovered four main benefits when working with triples. First, triples can be more expressive and less ambiguous descriptions of abstracted tasks compared to single-terms and tuples. Second, when several (multimodal) types of data are involved, the distinction between different criteria can be particularly useful. Third, the matrix notation is useful to validate and assess the diversity of tasks, as similar tasks are likely to have similar matrix signatures. Finally, triples provide a clear recipe that can be used by designers and developers of interactive visual data analysis tools.

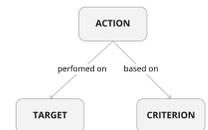

We suggest a matrix representation to illustrate interactions between targets and criteria, as the wrap figure shows. A diagonal pattern in the matrix indicates the traditional usage of targets and implicit criteria, in which no distinction is made between the two, such as: group (action) TSES (target, criterion). Any other cell in the matrix illustrates the added expressiveness of the triple notation: A type of action can be performed on *m* targets while being based on *n* criteria, even in the same visualization approach, such as: group (action) TSES (target) by features (criterion) or by metadata attribute (other criterion). Using this matrix representation, researchers can obtain a precise description of users' tasks which can guide the visualization tool design. We will make use of triples when applying our methodology to TSES in Section 5. In our typology in Table 2, we demonstrate the matrix notation for all 510 coded triples from both study sources.

## 5 APPLICATION OF THE METHODOLOGY

We apply our novel methodology to TSES, following two overarching goals. First, we will validate the methodology by demonstrating its usefulness for building dataset-centric taxonomic structures for tasks. Second, we aim to build the first task typology for TSES. We summarize the decisions and actions made during the five phases, Figure 3 summarizes our approach. Additionally, ten detailed figures in the supplemental materials further illustrate this process.

### 5.1 Study Data Collection

We performed qualitative analyses based on two complementing sources: a user survey study and a literature study. We opted for a larger-scale *user survey*, which leads to a great variety of analysis tasks that users are interested in, to avoid skewing our analyses towards a limited set of cases. We also decided on a *survey on design studies* to incorporate knowledge in visualization.

#### 5.1.1 User-Based Survey Study

Survey Creation: We conducted an anonymous online survey with qualitative and quantitative questions to individuals interested in analyzing TSES. From each response, we wanted to learn about the real-world scenario described, along with details about the domain context, data characteristics, user goals, data analysis tasks, current data analysis methods, and capabilities of available tools. To familiarize participants with the concept and definition of TSES and to highlight differences between classical event sequences, we used the real-world example of "traffic accidents" throughout the survey. Also, we included a series of illustrations along with the questions to enhance the comprehensibility.

Survey Distribution and Responses: We recruited participants in our academic environment. Participants were required to have at least an undergraduate degree, express an interest in TSES, and be proficient in English. We received 70 survey responses, spanning more than 25 domains, such as sports, accidents, food, technology, retail, health, and nature. The cases mentioned by participants were highly diverse, including examples such as 'eating doughnuts', 'food deliveries during lockdown', 'gaining Instagram followers', or 'purchasing airline tickets'.

#### 5.1.2 Survey on Design Studies

Paper Selection: As only a very limited number of design studies on TSES exist, we chose papers in the realm of other time-oriented data, while maximizing the variety across application domains and user groups. The paper selection process was:
1. *Preliminary Selection*: search for literature using keyword combinations such as 'event', 'sequence', 'data', 'time', 'temporal', 'time-stamped', or 'time-oriented', leading to 55 papers.
2. *Secondary Selection*: selection of a subset of design studies and survey papers with clear summaries of task abstractions, finalized requirements, or state-of-the-art analytical tasks.
3. *Final Selection*: in-depth examination of each paper and exclusion of papers with tasks that hide data-specific information (cf. Section 2.2), resulting in 14 design studies [17, 20, 31, 35, 47, 48, 50, 61, 62, 80, 86, 91, 93, 94] and 2 survey papers [33, 66].

### 5.2 Coding

Two authors were involved in the coding phase, either creating codes for extracted tasks, finding similar tasks for existing codes, or reviewing tasks that had already been coded. This pertained to both the user-based survey data and the design study survey data. Conflicting cases were resolved iteratively, sometimes with the help of a third party. While the two studies formed parallel processes, we did the coding interchangeably. Further, to code each study, we adopted the different processes described below. This resulted in two code sets, one for each study.

#### 5.2.1 User-Based Survey Study

The user survey responses provided interesting insights into individual real-world scenarios, domain and user contexts, user goals. Further, participants provided information about the TSES, useful for data characterization purposes. To code responses, we applied hierarchical task abstraction [93] to decompose domain-specific tasks into intermediate or lower-level tasks, and abstracted the tasks from domain-specific language. During the process, we identified five irrelevant responses which we removed from further analysis.

#### 5.2.2 Survey on Design Studies

For the design studies, we followed the open coding [83] approach. For every paper, our goal was a summary of abstracted tasks and requirements. In the event that the task summaries were too domain-specific, we added a step to convert tasks to a more abstract form.

### 5.3 Task Categorization

We applied affinity diagramming to form groups of tasks for the two individual sources. Again, two authors were involved in the iterative process. Codes within a group were reviewed by both authors, in case of conflicting or opposing opinions the discussion was opened to all authors. Later in the process, we applied split and merge operations on individual task groups as needed. With the aim of generalizability, we excluded every task mentioned only once (in one user survey or in one design study). As an intermediate result, the two authors identified 33 tasks from the user survey responses and 36 tasks from the design studies, followed by a discussion and refinement process among all authors leading to 25 user survey tasks and 26 design study tasks.

| Action | US | DS | Action | US | DS |
|---|---|---|---|---|---|
| Derive Metrics | 39 | 12 | Identify Motifs | 10 | 10 |
| Summarize | 32 | 14 | Add/Modify | 8 | 3 |
| Group | 29 | 11 | Filter | 7 | 9 |
| Compare | 26 | 12 | Analyze State Transition | 7 | 4 |
| Relate | 23 | 6 | Compare Threshold | 7 | 4 |
| Identify Common | 22 | 7 | Recommend | 7 | 4 |
| Analyze Trends | 21 | 3 | Gain Overview | 6 | 7 |
| Emphasize | 16 | 8 | Align | 6 | 3 |
| Annotate | 14 | 4 | Detect Outliers/Anomalies | 3 | 12 |
| Show Details | 13 | 14 | Analyze Fluctuations | 6 | - |
| Segment | 12 | 5 | Merge | 3 | - |
| Sort/Rank | 12 | 5 | Discover The Causality | - | 7 |
| Find Similar | 11 | 9 | Save Subset Of Interest | - | 3 |
| Predict | 11 | 2 | Identify Distinct Entities | - | 3 |

Table 1: Comparison between task occurrences in the two studies: the 65 user surveys (US, 351 codes) and the 16 design studies (DS, 181 codes). Most of the tasks for TSES are common in both studies.

### 5.4 Task Synthesis

We followed the strategy of *merging* the two task categorizations, as we had two equally strong sources of information. This approach included merging, splitting, and re-wording operations of tasks to further facilitate the unification. Table 1 provides summaries of task frequencies in both sources. More than 80% of tasks were identified by both sources (96% of the codes), two tasks occur only in the user survey, and three tasks occur only in the design studies. We chose to use only tasks that occur in both sources which is the most generalizable option, as stated in the methodology. This resulted in 23 tasks occurring in both sources, which we will describe in Section 6 in detail. Additionally, we will discuss the two user survey-only tasks, as they may signify a potential novel design opportunity.

### 5.5 Action-Target-Criteria Crosscut

In the final phase, we crosscut the actions of the 23 derived tasks with the data targets, i.e., with the data characteristics of TSES. Following our proposal of triples in Section 3.5 as a more fine-grained description of tasks, we performed an action-target-criteria crosscut. For TSES, targets and criteria can be based on five distinct data characteristics:
- **Events**: the most fine-grained existing data object
- **Event Sequences [ES]**: entire sequences of events, or subsets
- **Groups of Event Sequences [Gr(ES)]**: the result of ES grouping
- **Metadata Attributes**: data attributes for contextualization
- **Metrics/Features**: numerical attributes, metrics output

Note that metrics/features never form a target but are only leveraged as criteria, leading to four targets and five criteria. In the course of the process, we integrated other data-related terms found in the survey data such as "patterns", "trends", "motifs", and "time" into the set of actions and targets/criteria. The action-target-criteria crosscut for the 23 tasks leads to the typology presented in Table 2.

The procedure was as follows: for each of the 23 tasks, we traversed all associated codes in the affinity diagram. Whenever we identified a previously unseen combination of action, data target, and data criterion in one of the codes, we marked this in the table with •, and added the respective signature in the matrix mosaics ▦ with targets forming lines and criteria forming columns. Details about the resulting taxonomic structure are highlighted in Section 6.

## 6 TIME-STAMPED EVENT SEQUENCES: A TASK TYPOLOGY

We present a typology of tasks for TSES, shown in Table 2. The structure of the typology is a table, with the 23 *actions* of tasks as rows. The typology is special in that two data characteristics are included as the main columns of the table: *targets* and *criteria* (cf. Section 3.5). Using the "group event sequences by metadata attribute" as an example, this code is interpreted as an *action* (group), applied to a data *target* (event sequences), by a data *criterion* (metadata attribute). As an alternative, "group event sequences by features" also has the group action and the event sequences as data targets, but differs in the criterion that is applied (features). With this recipe, visualization designers exactly see, e.g., that a clustering method may be an appropriate design choice to support the latter task. The types of data *criteria* mainly align with data *targets*. Events, event sequences, groups of event sequences, and metadata pertain to both (shown as columns in Table 2). A unique data characteristic in the criteria is *feature*, to account for the many metrics that can be applied to TSES, allowing machine learning support but also interactivity such as rank-by-feature.

The Target × Criteria column shows that occur in the two sources examined. It can be seen that a variety of combinations of targets and criteria can be observed for each task. Using the *summarize* task as an example, 13 of the 4 × 5 possible combinations have been mentioned. We present an additional column in Table 2 that allows further classification of the 23 tasks. The *Terminal Task* criterion helps readers distinguish between task serving as a means, or as an end (terminal task). This gives guidance on possible higher-level tasks or task combinations in workflows, such as sort (T12) or filter (T17) actions to compare (T4) interesting TSES, or show details (T10). In Table 3, we describe each of the 23 tasks in detail, in the order of frequency of occurrence in the user survey study.

## 7 EVALUATION

We evaluate the typology by interviewing a domain expert about his dataset of TSES, and subsequently categorizing his tasks using the typology (see Table 1 in the supplemental materials). Andreas, a researcher with six years of experience in the field of cybersecurity threats, served as our domain expert. Two authors conducted an 65-minute video call interview. During the interview, we first asked the domain expert to describe the characteristics of his dataset, including information about the data semantics. In a second step, we asked Andreas to explain what his workflow would look like when analyzing this dataset with a powerful analytics tool, beyond the capabilities of his current solutions. His dataset is related to host communications in a computer network. He defines events as timestamps when a host machine communicates, while the temporal signature of the communication of a host would represent an event.

Andreas stated that when analyzing a new dataset, he would begin by looking at an overview of the data (T21) to detect prominent patterns (T15) across the dataset or across multiple hosts. If he finds a single or recurrent communication pattern of interest (T15), he will inspect (T10) the host communicating in that manner. Subsequently, he would try to compare the host to other hosts (T4), aiming for TSES with particularly high similarity or dissimilarity (T13), or motifs of particularly high regularity (T1). Andreas also mentions that he would label a communication pattern (T9) if it were especially noticeable. He would then try to find communication patterns that are similar (T13) to the labeled pattern and determine how many of these patterns occur in each host, e.g., by sorting (T12) them by the occurrence of the pattern for corresponding features (T1).

He would also cluster (T3) the dataset to uncover possible interesting communication patterns within a cluster of comparable (T15) hosts. He would then anticipate some sort of visualization of the clustering result, and a way to subsequently label (T9) interesting clusters. If he discovers a unique signature (T4, T15) in one of the clusters, he would strive for detail about this cluster (T10). Further, he wants to filter (T17) the dataset by these clusters so that only the data of these hosts remain in the analysis.

| Task ID | Action | Target | | | | Target × Criteria | Criteria | | | | | Terminal Task |
|---|---|---|---|---|---|---|---|---|---|---|---|---|
| | | *Event* | *ES* | *Gr(ES)* | *Metadata* | | *Event* | *ES* | *Gr(ES)* | *Metadata* | *Feature* | |
| T1 | Derive Metrics | • | • | • | | | • | • | • | • | | No |
| T2 | Summarize | | • | • | | | • | • | • | • | • | Yes/No |
| T3 | Group | | • | • | | | • | • | • | • | • | No |
| T4 | Compare | • | • | • | • | | • | • | • | • | • | Yes |
| T5 | Relate | • | • | | • | | • | • | • | • | • | Yes |
| T6 | Identify Common | | • | | • | | • | • | • | • | • | Yes/No |
| T7 | Analyze Trends | • | • | • | • | | • | • | • | • | • | Yes |
| T8 | Emphasize | • | • | • | | | • | • | • | • | • | Yes |
| T9 | Annotate | • | • | • | | | • | • | • | • | • | No |
| T10 | Show Details | • | • | | • | | • | • | • | • | • | Yes |
| T11 | Segment | | • | | | | • | • | | • | • | No |
| T12 | Sort/Rank | | • | | • | | • | • | • | • | • | No |
| T13 | Find Similar | • | • | | • | | • | • | • | • | • | No |
| T14 | Predict | • | • | | • | | • | • | • | • | • | Yes |
| T15 | Identify/Simplify Motifs | | • | | | | • | • | | | • | Yes/No |
| T16 | Add/Modify | • | • | | | | • | | | • | | No |
| T17 | Filter | • | • | • | | | • | • | | • | • | No |
| T18 | Analyze State Transition | • | • | | • | | • | • | | • | • | Yes |
| T19 | Compare Threshold | • | • | | • | | • | • | | • | • | Yes |
| T20 | Recommend | • | • | | • | | • | • | • | • | | No |
| T21 | Gain Overview | • | • | • | • | | • | • | • | • | • | No |
| T22 | Align | | • | | | | • | • | | • | | No |
| T23 | Detect Outliers/Anomalies | • | • | • | • | | • | • | • | • | • | Yes/No |

Table 2: Data-Centric Task Typology for TSES with 23 tasks. The actions of tasks can be applied on four data targets (event, event sequences (ES), groups of event sequences (Gr(ES)), and metadata). Five criteria can be used to perform actions on targets (event, event sequences (ES), groups of event sequences (Gr(ES)), metadata, and features). Example: group (action) event sequences (target) by metadata attribute (criterion). Target × Criteria summarizes all pairwise combinations mentioned in the two sources and can be used to form precise recipes for tool-building.

Continuing the idea of comparing different groups (T4), Andreas mentions that he would like to detect outliers (T23) in the dataset and then compare hosts (T4) that contain outliers to more ordinary sequences. By doing so, he aims to detect irregular network activity and potential threats. For the identification of irregular activity, he would, however, also need contextual information (T5, T10) about the host and the corresponding environment he is currently looking at to explain this irregular activity. Additionally, he would try to confirm the irregularity of outliers (T23) within a particular host by looking at the occurrence of outliers across different hosts (T1, T4).

## 8 DISCUSSION

**Actions, Targets, and Criteria:** We have introduced the notation of triples to describe tasks as an extension of actions/targets [59, 71] and targets/constraints [5, 12]. Beyond TSES, it will be interesting to investigate the usefulness of the triple notation for other data types, and for use by visualization designers [53] in general.

**Parallel Abstraction of Actions and Targets:** Our methodology presumes that abstractions for *data* already exist for the action-target crosscut phase. If researchers cannot benefit from pre-existing data characterizations, an explicit coding, categorization, and synthesis process along with the methodology would be an extension.

**Time Criterion:** We identified that time is often useful as a criterion to perform actions on targets of interest. Examples include *segmenting* or *aligning* sequences by time, or *filtering* sequences by time intervals. We have combined time with the event criterion, as they have the same granularity and a similar syntax. However, for some tasks it may also be useful to add time as a first-class citizen.

**Terminology: Pattern** In the typology, we avoid the use of the term pattern, which was prominently used in both sources. However, the notion of patterns was used in very diverse ways, similar to the concepts of *findings*, across all data targets and criteria. We distributed the usage of patterns across the four characterized data targets, with motifs being subsequence patterns in TSES.

**Excluded Tasks** Our merging strategy was geared towards the most generalizable option, i.e., all 23 tasks mentioned in both sources were used. However, the discrepancy between survey respondents and the literature survey led to removal of five other tasks, or 4% of all codes (cf. Table 2). We highlight the two tasks named only by survey respondents, i.e., tasks not yet abstracted in visualization research in the broader scope of temporal data. These tasks may signify a potential novel design opportunity. *Analyze Fluctuations* aims to identify changes in trends and metrics over time, *Merge* refers to merge actions on several segments/event sequences for downstream tasks like deriving metrics, or analyzing trends.

## 9 CONCLUSION

We proposed a methodology that combines user-based studies and literature-based studies to jointly build data-centric taxonomic structures for tasks. We applied the methodology to develop a task typology for TSES, a type of data that has been little investigated and for which the methodology can be leveraged to its full potential. A total of 65 survey responses and 16 design studies related to TSES formed the basis for the application of the five phases of the methodology, resulting in 23 clearly separable tasks. Building upon these tasks, we presented a data-centric task typology for TSES that uses triples for the description of tasks (actions, applied on targets (data), based on criteria (data) which appeared to be a useful language of description of tasks for TSES. We validated the typology in an evaluation study with a domain expert performing a real-world case on TSES in cybersecurity. With the methodology, the task typology, and the use of triples, we aim to inspire future work on taxonomy-building and on designing novel interactive visual analysis systems. Future work includes elaborating further on triples to describe data-centric tasks. Also, it will be interesting to experience other cases that can benefit from a methodology combining two complementary sources of study data.

| ID | Action | Targets and Criteria, all mentioned combinations of targets × criteria are shown in the matrix | Matrix |
|---|---|---|---|
| T1 | Derive Metrics | Derive a feature by applying carefully selected metrics to events, TSES, and groups of TSES (targets). Criteria for such metrics can be based on events, TSES, groups of TSES, and on metadata. Most common metrics requested by users include regularity, density, gaps, frequency, and event count. *Example code*: Derive metrics (action) at the granularity of events (target) based on metadata attribute (criterion). | 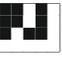 |
| T2 | Summarize | Summarize TSES at all three possible granularities (targets). Especially for large datasets, summaries bring important information upfront, while ignoring unnecessary details. The use of statistics of metadata as well as features (criteria) was frequently observed in the coding, shifting the analysis of TSES towards distributions. *Example code*: Show summary (action) of groups of TSES (target) by the event regularity feature (criteria). | 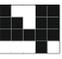 |
| T3 | Group | Grouping of events, TSES, and groups of TSES (targets) by events, TSES, metadata, or features (criteria). Clustering methods play an important role when targets are grouped (action) by features of the TSES content, or by metadata. *Example code*: Group (action) TSES (target) by a metadata category (criterion). | 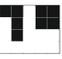 |
| T4 | Compare | Compare events, TSES, groups of TSES, or metadata (targets) to identify commonalities and differences. In general, comparison is typically applied between items of a similar type (a TSES can hardly be compared to a metadata attribute). *Example code*: Compare (action) a set of TSES (target) by motif patterns (criterion). | 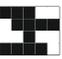 |
| T5 | Relate | Relate events, TSES, or metadata (target) to another criterion (events, TSES, groups of TSES, metadata, and feature). Relate is a counterpart to compare: it is typically applied across data characteristics, i.e., some identified phenomenon for a target can be contextualized by a criterion. *Example code*: Relate (action) group of servers TSES (target) with the host metadata attribute (criterion), revealing that all TSES are from the same host. | 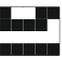 |
| T6 | Identify Common | Identify (most) common occurrences of TSES or metadata (targets). Criteria are all three granularities of TSES, metadata, as well as features. In many cases, users even asked for *most* common occurrences, which amplifies to focus on predominantly occurring phenomena in the data. *Example code*: Identify (action) the most occured metadata value (target) of all TSES (criterion). | 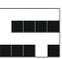 |
| T7 | Analyze Trends | Analyze trends within events, TSES, groups of TSES, or metadata (targets) based on all five types of (criteria). Users are particularly interested in analyzing daily, monthly, and other seasonal trend criteria in TSES, all of which are examples of trends based on temporal metadata attributes. *Example code*: Analyze trends (action) for each TSES (target) based on the acceleration metrics (criterion) | 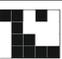 |
| T8 | Emphasize | Emphasize events, TSES, or a group of TSES (targets) of interest that helps users to make interesting findings. We identified all five types of criteria for emphasize tasks. *Example code*: Emphasize (action) a trend in TSES (target), using the acceleration feature (criterion). | 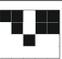 |
| T9 | Annotate | Annotate an event, TSES, or a group of TSES (targets) with additional information, such as a descriptive label. Among the user surveys, annotation differs between manually added annotations by humans and (semi-) supervised annotation with computational annotation support. All five types of criteria can be useful for annotate tasks. *Example code*: Annotate (action) TSES (target) based on set of metadata (criterion). | 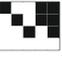 |
| T10 | Show Details | Show details on demand for specific events, TSES, groups of TSES, or metadata (targets). Criteria to be shown are events, TSES, groups of TSES, and metadata. *Example code*: Show details (action) of events (target) using metadata attributes (criterion). | 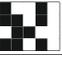 |
| T11 | Segment | Segment actions are explicitly applied at the granularity of TSES (targets), to partition TSES into more manageable and useful temporal segments. While users are interested in splitting criteria based on temporal events and temporal metadata such as day, week, month, features form another important segmentation criterion. *Example code*: segment (action) TSES (target) by a certain date (criterion). | 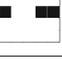 |
| T12 | Sort/Rank | Sort representations of TSES or metadata (targets). Sorting criteria can be of all five types, with an observed dominance of features (sort-by-feature), or metadata attributes. Sorting allows users to faster identify interesting TSES for detailed analysis, while discovering new findings due to the different ordering of TSES. *Example code*: Sort (action) TSES (target) by the number of events (criterion). | 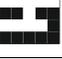 |
| T13 | Find Similar | Find similar events, TSES, or metadata (targets) based on selecting or formulating a query, respectively. Observed criteria are events, TSES, groups of TSES, metadata, and predominantly: features, in line with information retrieval methods. *Example code*: Find similar (action) TSES (target), based on the similarity notion determined by a set of features (criterion). | 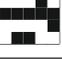 |
| T14 | Predict | Predict an event, TSES, or metadata (targets), by criteria such as events, TSES, groups of TSES, metadata, or features. For TSES, we identify two dominating types: temporal prediction (the likelihood of an event) and value prediction (values for metadata attributes, e.g., for missing value imputation). *Example code*: Predict (action) the temporal occurrence of an earthquake event (target) by taking historic TSES data into account (criterion). | 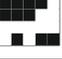 |
| T15 | Identify/ Simplify Motifs | Identify and simplify motif patterns in TSES (targets) that stand out, based on events, TSES, or groups of TSES. Simplification refers to the idea to substitute/label motif occurrences by a (visual) placeholder and is useful when complex datasets are to be simplified, or when supervised motif detection methods are to be trained. *Example code*: Identify patterns (action) of TSES (target) based on event signatures (criterion). | 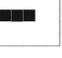 |
| T16 | Add/ Modify | Add or modify a single event or TSES (targets) based on temporal event or metadata criteria. Add/modify can help to impute missing values, improve the data quality, or adapt the data towards the user's information need. *Example code*: Add (action) missing event (target) based on event context (criterion). | 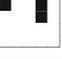 |
| T17 | Filter | Filter out irrelevant events, TSES, or groups of TSES (targets) to keep only the relevant. Users can filter by metadata or by feature criteria, events, or the time period of TSES. *Example code*: filter (action) TSES (target) by the regularity feature (criterion). | 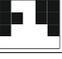 |
| T18 | Analyze State Transition | Analyze the transitions of states of events, TSES, or metadata (targets) changing over time. State transitions can be analyzed based on events, TSES, metadata attributes, and features. *Example code*: Analyze (action) the transition of event states (target) with respect to environment parameters covered through metadata (criteria). | 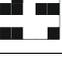 |
| T19 | Compare Threshold | Compare the threshold of an observed TSES or metadata attribute (target) with an expected value in the metadata or the features (criteria). In many cases, users have clear expectations for TSES when it comes to values of similarities, metadata, and features. *Example code*: Compare (action) the number of expected traffic jams with the observed number (criterion) for a city (TSES, target). | 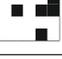 |
| T20 | Recommend | Recommend reliable suggestions for events, TSES, or metadata (target), based on the current state of analysis. Criteria include events and corresponding time information, TSES, groups of TSES, as well as metadata. *Example code*: An event (target) is recommended (action) by the system based on a performance feature (criterion). Such a performance feature could be the rainfall, which may trigger decisions on sewage systems. | 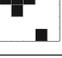 |
| T21 | Gain Overview | Gain an overview of all events, TSES, groups of TSES, or metadata (targets) based on events, TSES, groups of TSES, and metadata (criteria). An overview is useful to identify patterns more easily, such as patterns/motifs and dense/sparse regions. Overview differs from summarize in that all data is shown, not only summarized abstractions. *Example code*: Gain overview (action) of TSES (target) structured by events/time and event count (criteria). | 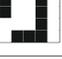 |
| T22 | Align | Alignment eases sequence comparison, with TSES being the only target mentioned. The dominating criterion is time points of events, as an alternative temporal metadata attributes can determine TSES alignment. Alignment eases the comparison of TSES, to better assess seasonality, or to synchronize TSES by a particular point in time. *Example code*: Align (action) TSES (target) by a particular date event, such as September 11, 2001 (criterion). | 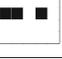 |
| T23 | Detect Outliers/ Anomalies | Detect outliers or anomalies from an event distribution, TSES, groups of TSES, or metadata (targets). Criteria can be based on events, TSES, groups of TSES, metadata, and features in particular. According to the user survey, outliers tend to be more extreme values, whereas anomalies are rather unexpected or scarce, often with a strong semantic impact. *Example code*: Detect (action) outlier of an implausibly high count of events (criterion) in a TSES (target). *Example code*: Detect (action) fraud anomaly (criterion) in a network communication TSES (target). | 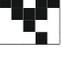 |

Table 3: Detailed description of actions, targets, and criteria for all 23 tasks. Matrices show all mentioned targets × criteria combinations.